\documentclass[a4paper,fleqn,usenatbib]{mnras}
\usepackage{newtxtext,newtxmath}
\usepackage[T1]{fontenc}
\usepackage{ae,aecompl}
\usepackage{graphicx}	
\usepackage{amsmath}	
\usepackage{amssymb}	
\usepackage{braket}
\usepackage{siunitx}
\usepackage{mathtools}
\usepackage{flexisym}
\usepackage{hyperref}
\usepackage{mathptmx}
\usepackage{flushend}

\title[The use of OPA in astronomy]{The usability of optical parametric amplification (OPA) of light for high angular resolution imaging and fast astrometry}
\author[A. R. Kurek et al.]{
	A.~R. Kurek,$^{1}$\thanks{E-mail: aleksander.kurek@uj.edu.pl (ARK)}
	A. Stachowski,$^{1}$
	K. Banaszek$^{2, 3}$
	and A. Pollo$^{1, 4}$
	\\
	$^{1}$Astronomical Observatory of the Jagiellonian University, Orla 171, 30-244 Cracow, Poland\\
	$^{2}$Faculty of Physics, University of Warsaw, Pasteura 5, 02-093 Warsaw, Poland\\
	$^{3}$Centre of New Technologies, University of Warsaw, Banacha 2c, 02-097 Warsaw, Poland\\
	$^{4}$National Centre for  Nuclear Research, Ho\.za 69, 00-681 Warsaw, Poland
}

\date{
MNRAS, vol. 476, iss. 2, p. 1696 -- 1704\quad(11 May 2018)\\
DOI: \href{https://doi.org/10.1093/mnras/sty307}{10.1093/mnras/sty307}\\
Accepted:  1-Feb-2018. Received:  11-Nov-2017
}

\pubyear{2018}

\begin{document}
\label{firstpage}
\pagerange{\pageref{firstpage}--\pageref{lastpage}}
\maketitle

\begin{abstract}
High angular resolution imaging is crucial for many applications in modern astronomy and astrophysics. The fundamental diffraction limit constrains the resolving power of both ground-based and spaceborne telescopes. The recent idea of a Quantum Telescope based on the Optical Parametric Amplification (OPA) of the light is aimed at bypassing this limit for imaging of extended sources by an order of magnitude or more.\\
We present an updated scheme of an OPA-based device and a more accurate model of signal amplification by such device. A semiclassical model we present predicts that the noise in such a system will form so called light speckles due to the light interference in the optical path. Based on this model, we analyzed the efficiency of the OPA in increasing the angular resolution of imaging of extended targets and precise localization of a distant point source.\\
According to our new model, OPA is offering a gain in resolved imaging in comparison to classical optics. In the same time, we found that OPA can be more efficient in localizing a single distant point source in comparison to classical telescopes.
\\
\end{abstract}

\begin{keywords}
telescopes -- instrumentation: high angular resolution -- techniques: high angular resolution -- astrometry
\end{keywords}

\section{Introduction}
Present technology does not allow for precise resolved imaging of a vast majority of astrophysical targets of interest due to the large distances. Current technological limits of the diameter of the telescope primary mirror are: $\sim$40\,m on the ground~\citep{EELT1, EELT2} and $\sim$6.5\,m in the space~\citep{JWST}. This is still many orders of magnitude below the size which allows for resolved imaging of such celestial objects like most of stars or central parts of galaxies. Therefore, the brightness, temporal brightness variations (lightcurves) and temperatures are the primary source of information about the physical sizes of observed objects. The highest angular resolution available today in the optical range of wavelength (which in astronomy is usually considered as \mbox{400\,--\,1100\,nm}) is $\sim$0.1\,arcsec, while the largest angular diameter of a star as seen from the Earth is 0.05\,arcsec (50\,mas; Betelgeuse~\citep{Betelgeuse}). Other desired targets are even much smaller angullary, e.g. the predicted size of prominent features of the event horizon of a black hole Sagittarius A* at the Galactic Centre is $\sim$30\,{\textmu}as~\citep{SaggitariusA}. Although existing optical interferometers achieve a resolution in imaging of up to 200\,{\textmu}as (\href{http://www.chara.gsu.edu/}{CHARA Interferometer}), their imaging capability is very limited and they can operate only on very bright targets~\citep{gravity}. The upcoming \mbox{\href{http://www.eventhorizontelescope.org/}{Event Horizon Telescope}} is expected to provide polarimetric imaging with \,25\,{\textmu}as resolution at 1.3\,mm in 2018~\citep{EHT}. But no existing telescope project is aimed at improving the diffraction limit of the instrument~\citep{instrumentsReview}.

In order to increase the angular resolution, it is desired to read the information on the exact direction of every photon, which will contribute to the image, before most of this information will be disturbed due to diffraction in the optical system. This could be done if there existed a method to produce exact copies of each photon before the diffraction occurs, and to register the copies so that the statistical analysis can be performed (online or offline). In other words, it is desired to amplify the signal before registering it. Optical Parametric Amplification of the light (OPA) is a process where an input signal (photon) is amplified and its almost exact copies of it are created, at the expense of energy pumped in the amplifying medium (e.g. laser crystal). \cite{Qlimits} showed that this process is too noisy to immediately enable the increase of the resolution of imaging beyond the diffraction limit. However, in the recent years this limitation was reformulated: it was found that it is possible to decrease the registered noise amount significantly by the use of a trigger signal \mbox{\citep{noiseless1, cloning3}}. This idea of the OPA-based device for astronomical observations, often referred to as the Quantum Telescope (QT), was later further investigated by \cite{QT2} and \cite{QT3}.

In Sec.~\ref{sec:QTgeneral} of this paper we discuss the idea  of the telescope based on the OPA in detail. However, one drawback of all the models of the QT discussed so far was that they were based on very simplified models of noise. In the same time, the proper noise modeling in such systems, and especially
its statistical distribution, will be crucial for a proper assessment of the efficiency of such a device. As a step to address this issue, in this article we present a semiclassical model of the OPA as applied for astronomical imaging, which, in particular, provides a more accurate prediction of the noise than the models used before \citep{QT1, QT2, QT3}.

\vspace{2mm}
The article is organized as follows. In Sec.~\ref{sec:concept} we describe the idea of using OPA for increasing the angular resolution in astronomy. In Sec.~\ref{sec:model} we present previous research on the use of OPA in astronomy. In Sec.~\ref{sec:noise} we describe the influence of intrinsic OPA noise on the resolution gain. In Sec.~\ref{sec:newModel} we introduce semiclassical model of the OPA which we then apply for the efficiency characterization. Sec.~\ref{sec:sim} describes numerical simulations we performed. We use two methods of signal analysis: i)\,estimation of the position of the centroid of the signal and ii)\,estimation of the position of maxima of speckles. For the first method, results of the simulations of the efficiency in the angular resolution and localization of a distant point source are presented in, respectively, Sec.~\ref{sec:angRes} and~\ref{sec:loc}. For the letter method, the results are presented in Sec.~\ref{sec:REVangRes} and~\ref{sec:REVloc}. We conclude in Sec.~\ref{sec:Conclusions}.

\section{Quantum Telescope}
\label{sec:QTgeneral}
\subsection{Concept}
\label{sec:concept}
\begin{figure}
	\includegraphics[width=\linewidth, trim=0.7cm 0.8cm 1.7cm 0.3cm]{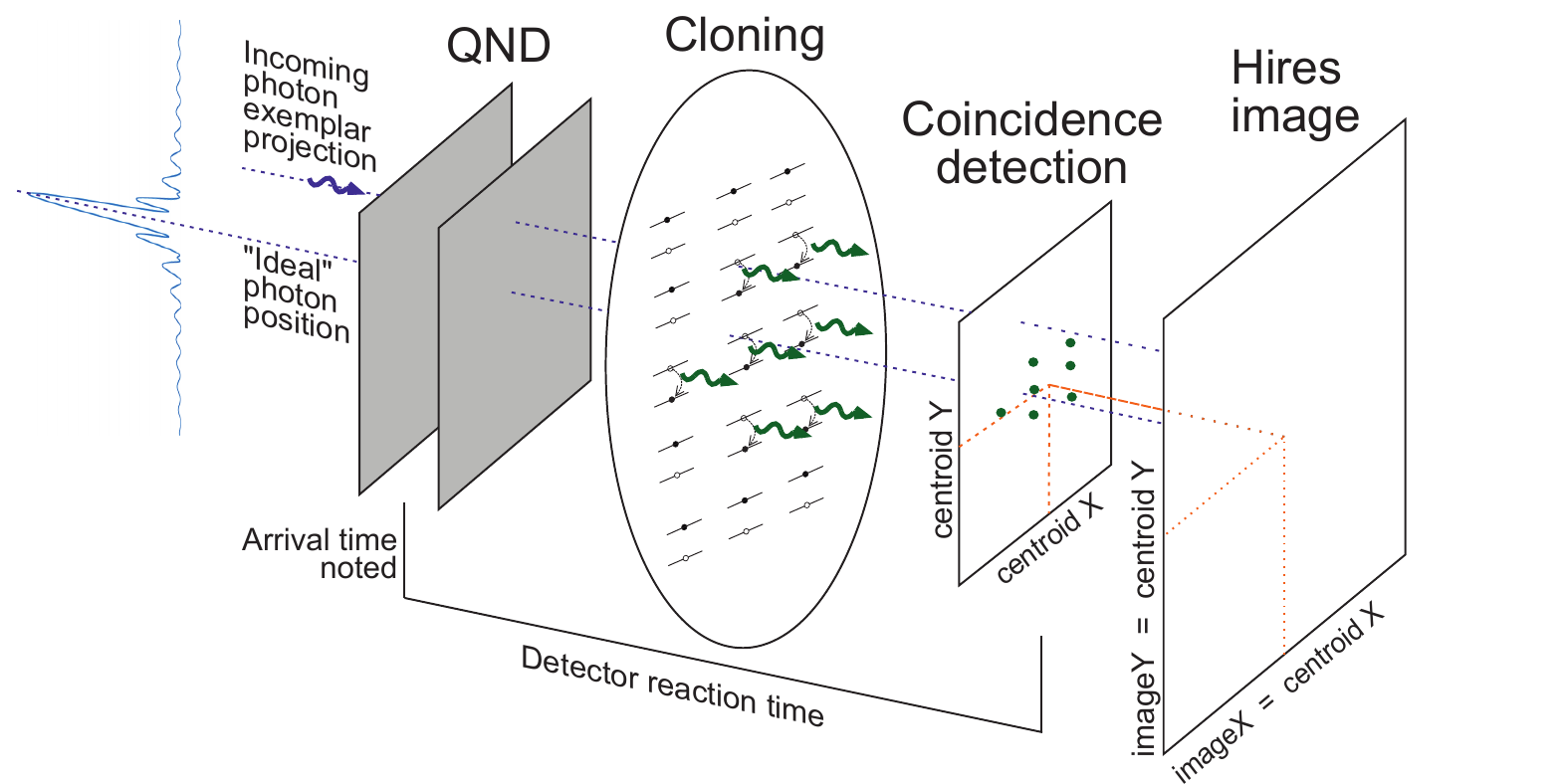}   
	\caption{
A general Quantum Telescope scheme. Elements are not to scale.
	}
	\label{fig:Idea}
\end{figure}

Recently, the idea of a Quantum Telescope (QT) was presented in~\cite{QT1, QT2}. The aim of the idea is to increase the resolving power of existing and future telescopes by introducing new optical elements into their optical path. In such a device each photon inbounding from a celestial extended source is first detected by the Quantum Non-Demolition (QND) device wherein the time of its arrival is registered. Quantum Non-Demolition (QND) is a type of measurement of a quantum system in which the observable is negligibly changed; see~\cite{QND} for further details. Most importantly, the photon is not absorbed by QND detection. As the arrival time is known, it is possible to turn on the detector only for a specified time interval, in which photon's arrival is expected. After the detection, the photon passes through a pumped amplifying medium (e.g. Beta barium BOrate crystal, BBO) and is cloned by parametric amplification~\citep{cloning1, cloning2, cloning3}. The photon from the astronomical target and its clones (stimulated emission) are from this moment indistinguishable and treated as one photon cloud which is registered by a fast two dimensional coincidence detector, e.g. ICCD (Intensified Charge-Coupled Device,~\cite{ICCD}) or EMCCD (Electron Multiplying Charge Coupled Device, also known as Low Light Level CCD, L3CCD) camera~\citep{emccd}. Unfortunately, the amplifier produces unavoidable spontaneous emission, which is required by the uncertainty principle~\citep{noiseReview}. Although very fast electronic gating of the detector (preferably not less than the coherence time of photons: $\Delta t = \lambda^2/(c\,\Delta \lambda)$) prevents the system from registering too much spontaneous emission photons, the final image is still contaminated by significant noise. Centroid position of the photons' cloud is computed offline and passed as a count into a final high resolution image. The entire process is illustrated in Fig.~\ref{fig:Idea}, which shows a toy-model of the QT. The process is repeated for every photon detected by QND and in this way the high resolution image is constructed, photon by photon, during sufficiently long exposition.

\subsection{Model}
\label{sec:model}
The first QT setup proposed in ~\cite{QT2} was based on a QND device with optical resonator (cavity). However, this would cause a large diffraction and prevent this QT setup from increasing the angular resolution. In~\cite{QT3} we suggested using the cavity-free QND device according to the scheme proposed by~\cite{QNDnoCavity}, which additionally is much easier to use than a "classical" QND proposed by~\cite{QNDnature}. In the cavity-free version the trajectory of a photon is not affected and the introduced diffraction is negligible, since a 1D waveguide embedded with atoms is used. In such a 1D waveguide (a hollow-core photonic crystal fiber), the light is well confined~\citep{confLight} in the 1D space as a guided mode (Keyu Xia, priv. communication).

\begin{figure}
	\includegraphics[width=\linewidth, trim=0cm 0.5cm 1cm 1.5cm]{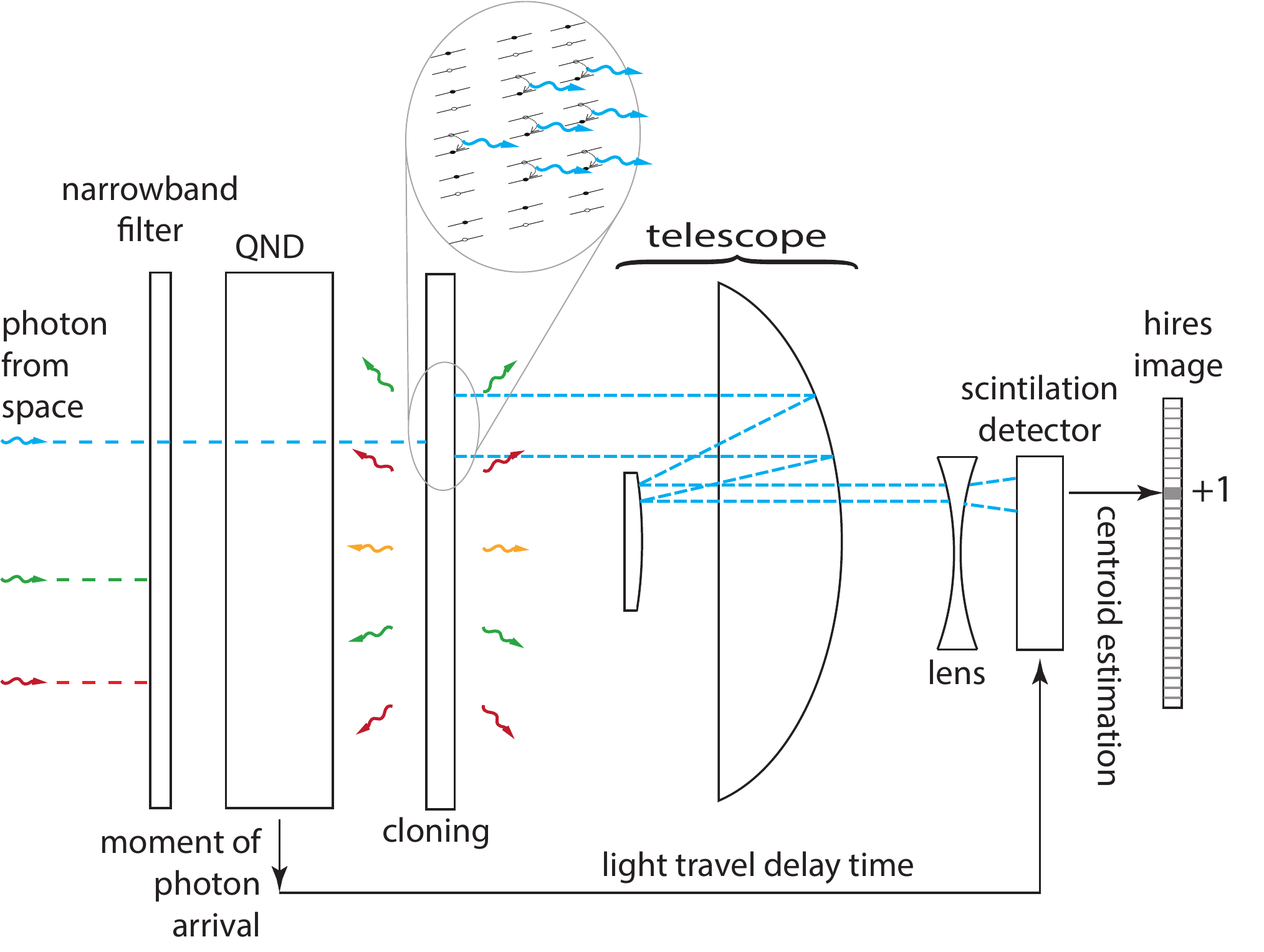}   
	\caption{
Scheme of QT. Quantum addition to the telescope is placed in front of the mirror.
	}
	\label{fig:bigQT}
\end{figure}

Additionally, as we discuss below, QT cannot be placed in the plane conjugated to the pupil, that is -- QT cannot be implemented as a small device located {\bf after} the mirror, as proposed by~\cite{QT4}. The basic theory of quantum physics and optics indicate that the position of a photon, whose wevefunction is reduced to the size of telescope`s aperture, cannot be retrieved with the accuracy exceeding the diffraction limit. Therefore, to extend the resolution of a telescope beyond this limit, the QND and cloning stages need to be placed {\bf before the aperture} (i.e. before the diffraction takes place), as shown in Fig.~\ref{fig:bigQT}. In this setup, the clones preserve the information of the original photon. The cloud of clones, although diffracted by the telescope aperture, still centres around the original position. In other words, each clone is diffracted and its position is randomly changed, but the cloud of clones preserves its centroid. Importantly, the detection in QND does not introduce diffraction, since only the arrival time of photon is detected, not its position.
In contrast, locating QND and cloning stages {\bf after} the telescope mirror cannot increase the resolution. The cloud of clones, as a whole, is randomly shifted, since the primary photon suffers from the diffraction. The QT placed behind the aperture works in fact as a simple photon intensifier (like a multichannel plate), which can never increase the resolution (usually degrading it). Design according to our proposed scheme (e.g.~Fig.~\ref{fig:bigQT}) is obviously more complicated technologically, since it requires a large amplifying medium and large QND detector. The QND are still in their initial stage of development, but, to our best knowledge, there is no premise for size limitations of such a device.

\subsection{Noise}
\label{sec:noise}
According to~\cite{errata} and \cite{Prasad}, QT has a large intrinsic noise: the amplification reduces the Signal-to-Noise Ratio (SNR) to $\lesssim$1/7.3. This was supposed to be the lowest fundamental limit of noise in this design, originating from the necessity of opening the wells of the detector for at least photon coherence time --- that is long enough for the photon to be absorbed~\citep{errata, Prasad}. Using Matched Filtering (MF) approach\footnote{The MF is an efficient detector of the known template in noisy environment and is widely used in radars or sonars, where the weak, well-defined reflected signals have to be detected~\citep{MF1, MF2}.} we proved~\citep{QT3} that even such a high level of noise does not prevent the QT from achieving a good performance, given that a sufficient number of clones is provided. QT toy-model was demonstrated to perform better than the Classic Telescope (CT) of the same mirror size, if on average $\geq$56 clones  are provided for every photon from space~\citep{QT3}.

It is obviously desirable to amplify a usually weak signal in astronomy. However, deterministic optical amplification (frequently referred to as quantum cloning of the light) has to be noisy~\mbox{\citep{Qlimits}}: any amplifier working independently on the phase of the input signal has to introduce noise because of the uncertainty principle and no-cloning theorem~\citep{noiselessAndScissors}. The desired perfect amplification for the QT would be noiseless, i.e. $\Ket{\alpha} \rightarrow \Ket{\sqrt{G}\alpha}$, where $\Ket{\alpha}$ is a coherent state of light and $G$ is an amplification gain. Such a procedure is also in principle possible probabilistically, given that one uses much more complex amplifier. A few working schemes were already demonstrated:~\cite{noiseless1, noiseless2, noiseless3}. However, all of them are very difficult to realize using the present technology, mainly because they use multiple quantum-scissors --- a device able to generate any desired superposition of the vacuum and one-photon states \citep{Qscissors1}. Nevertheless, functioning of such quantum-scissors was recently demonstrated by~\cite{Qscissors2, noiselessAndScissors}.
\vspace{2mm}

\section{QT\, --\, a semiclassical model}
\label{sec:newModel}
In the previous analysis~\citep{QT3} the noise was simulated by a Poissonian process and the clone count per frame was also distributed according to the Poissonian process. This is a default choice for rare events; see \citet[Chapt. 5]{qoBook} for derivation. A clones cloud was assumed to have the Gaussian shape. Below we present a more accurate 2D model of the QT that fully includes spatial intensity correlations in the image plane.
\vspace{2mm}

Let us consider first a point source located on the instrument axis. The normalized mode function in the pupil plane is defined over the circle $\Pi(\rho)$ of radius $D/2$. The circle has a following form:
\begin{equation}
	\Pi({\boldsymbol{\rho}}) = \begin{cases} 1/\sqrt{\pi D^2/4}, &  \quad \mbox{if $|{\boldsymbol{\rho}}| < D/2$} \\  
	0 & \quad \mbox{if $|{\boldsymbol{\rho}}| \ge D/2$}, \end{cases}
\end{equation}
where $D$ is the telescope mirror diameter. We use polar coordinates $\rho$ (radial) and $\phi$ (angular) in the pupil plane. The centre of the polar coordinate system is equivalent of the centre of the pupil plane. In order to account for the noise generated by the amplifier (``cloning device'') in the pupil plane, it is necessary to complement the signal mode\ \ $\tilde{u}_{00}(\rho,\phi)$ with orthonormal modes that form together a complete set over the pupil area. The signal can be decomposed by orthogonal modes in many different ways, so that we can choose any orthogonal set of functions for the description of the modes. Since we assume that the pupil is circular, the natural choice of orthogonal functions are the Zernike polynomials. A convenient choice~\citep{zernike} is given by the Zernike modes $\tilde{u}_{nm}({\boldsymbol\rho,\phi})$, where $n=0,1,2,\ldots$ and $m = -n, -n+2, \ldots, n-2, n$. Their explicit form can be found e.g. in~\cite{noll}.

In the image plane, contributions from individual modes are given by the corresponding two-dimensional Fourier transforms which explicitly read:
\begin{equation}
	u_{nm}(\tilde{\rho},\tilde{\phi})=(-1)^{n}\sqrt{n+1}\frac{J_{n+1}(2\pi\tilde{\rho})}{\pi\tilde{\rho}}
	\begin{cases}
		\begin{array}{l} \sqrt{2}\cos(\mid m\mid\tilde{\phi})\text{ if }m>0\\
			\sqrt{2}\sin(\mid m\mid\tilde{\phi})\text{ if }m<0\\
			1\text{ if }m=0 ,
		\end{array}
	\end{cases}
	\label{Eq:ModesImagePlane}
\end{equation}
where $J_n$ is the $n$-th Bessel function. $\tilde{\rho}$ and $\tilde{\phi}$ are Fourier-transformed variables. Units of $\tilde\rho$ are expressed in pixels.

We will assume that the amplifier has a uniform gain factor $G$ for all the spatial modes in the pupil plane, i.e. the amplitude of an individual mode is multiplied by a factor $\sqrt{G}$, whereas the mean number of spontaneously generated photons in that mode is $G-1$~\cite[Chapt. 9.3]{Haus2000}. We use a semiclassical model to describe light fluctuations, assigning to each mode a dimensionless complex amplitude $\alpha_{nm}$ whose squared absolute value $|\alpha_{nm}|^2$ characterizes the average photon number in that mode. In this paper, semiclassical model means that the model based on the first quantization theory: we treat the light beam as a classical electromagnetic wave and the amplification of the signal as a classical process, and we assume the semiclassical theory of photodetection. We assume the source is at the infinite distance. This assumption allows us to treat the signal as a coherent light with a unit amplitude. A given cloning medium is able to produce clones only in a very narrow wavelength span. So we can filter the light and pass only a given narrow wavelength span. Therefore, we can assume, that the photons' distribution is not thermal, and in consequence the difference in their wavelength is very small. A single-mode thermal light is governed by Bose-Einstein distribution~\citep{qoBook}. Because every emission in the cloning medium is independent of other ones~\cite[Chapt. 9]{Haus2000}, the singular emission is described by a singular mode. In consequence, a multi-mode distribution of the light is a result of convolution of a large number of single-mode distributions. Therefore, after the amplification process the amplitude is described by a probability distribution:
\begin{equation}
	p(\tilde{\alpha}_{00}) = \frac{1}{\pi(G-1)} \exp\left( - \frac{|\tilde{\alpha}_{00}|^2}{G-1} \right),
	\label{Eq:palpha00}
\end{equation}
where $\tilde{\alpha}_{00}=\alpha_{00}-\sqrt{G}$. The amplitudes $\alpha_{nm}$ for all other noise modes $u_{nm}$ with $n>0$ are characterized by a Gaussian thermal distribution:
\begin{equation}
	p(\alpha_{nm}) = \frac{1}{\pi(G-1)} \exp\left( - \frac{|\alpha_{nm} |^2}{G-1} \right), \quad n=1,2,\ldots \ .
	\label{Eq:palphanm}
\end{equation}
We use Eqs. (\ref{Eq:palpha00}) and (\ref{Eq:palphanm}) to find the modulus of the random coefficients $\alpha_{nm}$. Because Eq.~\ref{Eq:palphanm} is depended only on $|\alpha_{nm}|$, all the values of the phases of $\alpha_{nm}$ coefficients are treated the same by Eq.~\ref{Eq:palphanm}. In consequence, the values of the phases of $\alpha_{nm}$ coefficients are generated from the uniform distribution from the interval $\langle 0, 2\pi)$. If we know the value of the phase and the modulus of $\tilde{\alpha}_{00}$ coefficient, we know $\alpha_{00}$ coefficient because $\alpha_{00}=\tilde{\alpha}_{00}+\sqrt{G}$.

\begin{figure}
\centering
	\includegraphics[width=\linewidth, trim= 1.5cm 12.4cm 1cm 12.5cm]{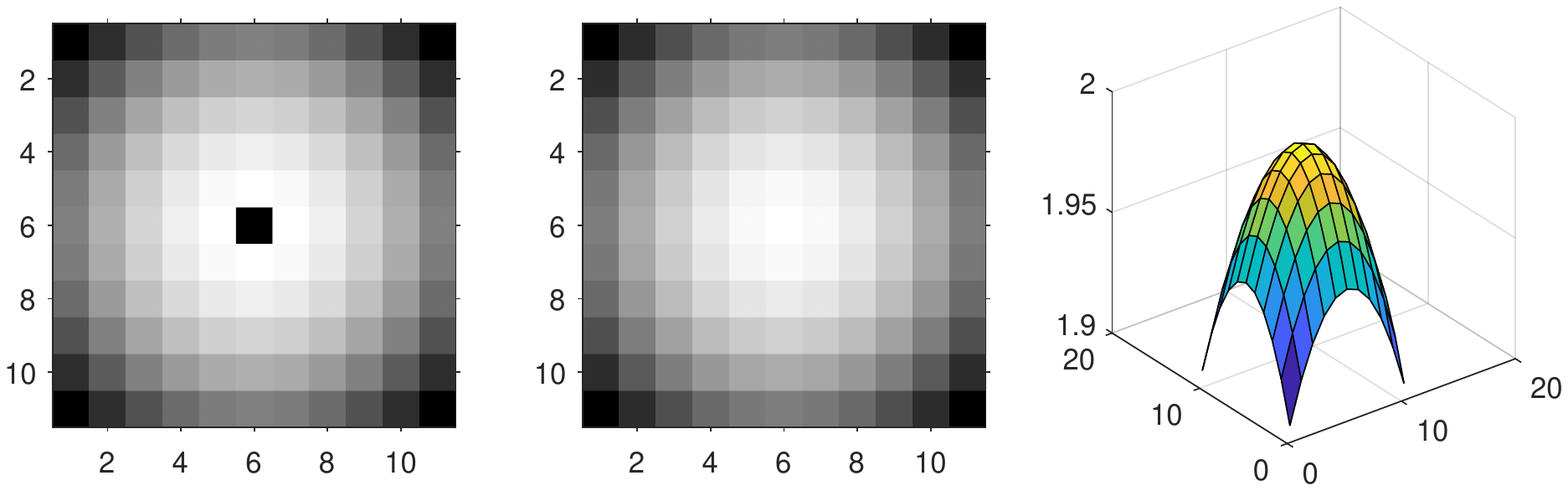}   
	\caption{
Electromagnetic wave density in the surrounding of $u_{1, 1}$(0,\,0). Left: pixel of undefined value visible in the middle. There are 4 highest points in this image: (5,\,6), (7,\,6), (6,\,5) and (6,\,7). Middle and right: value estimated from the fitting was pasted to this pixel. There is only one highest point: (6,\,6).
	}
	\label{fig:fitted}
\end{figure}

\begin{figure}
\centering
	\includegraphics[width=0.7\linewidth, trim= 4.5cm 7.2cm 2.8cm 4cm]{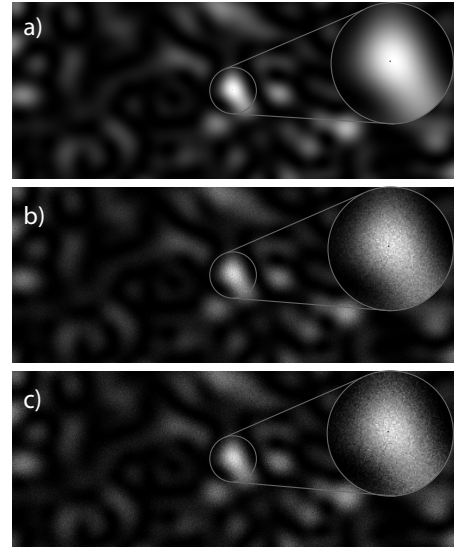}   
	\caption{
a) Exemplary realization of the density of the electromagnetic wave in the image plane for OPA gain $G$ = 15;\ b) Density of photon-events in the image plane;\ c) added simulated noise from high-end EMCCD / L3CCD camera image (gain\,=\,100, saturation level\,=\,10$^{\text{5}}$\,e$^{\text{-}}$). Realistic detector introduces also an additional uncertainty, mainly photon multiplication noise and Clock Inducted Charge (CIC) noise, which is omitted here, since in modern EMCCDs it is negligible. The scale depends on the distance of the image plane from the focus, therefore we do not associate it with a metric unit here. The black pixel inside the inset denotes the centre of the image.
	}
	\label{fig:images}
\end{figure}

\section{Simulations}
\label{sec:sim}

\begin{figure}
\centering
	\includegraphics[width=\linewidth, trim= 1.7cm 1.5cm 1.4cm 2cm]{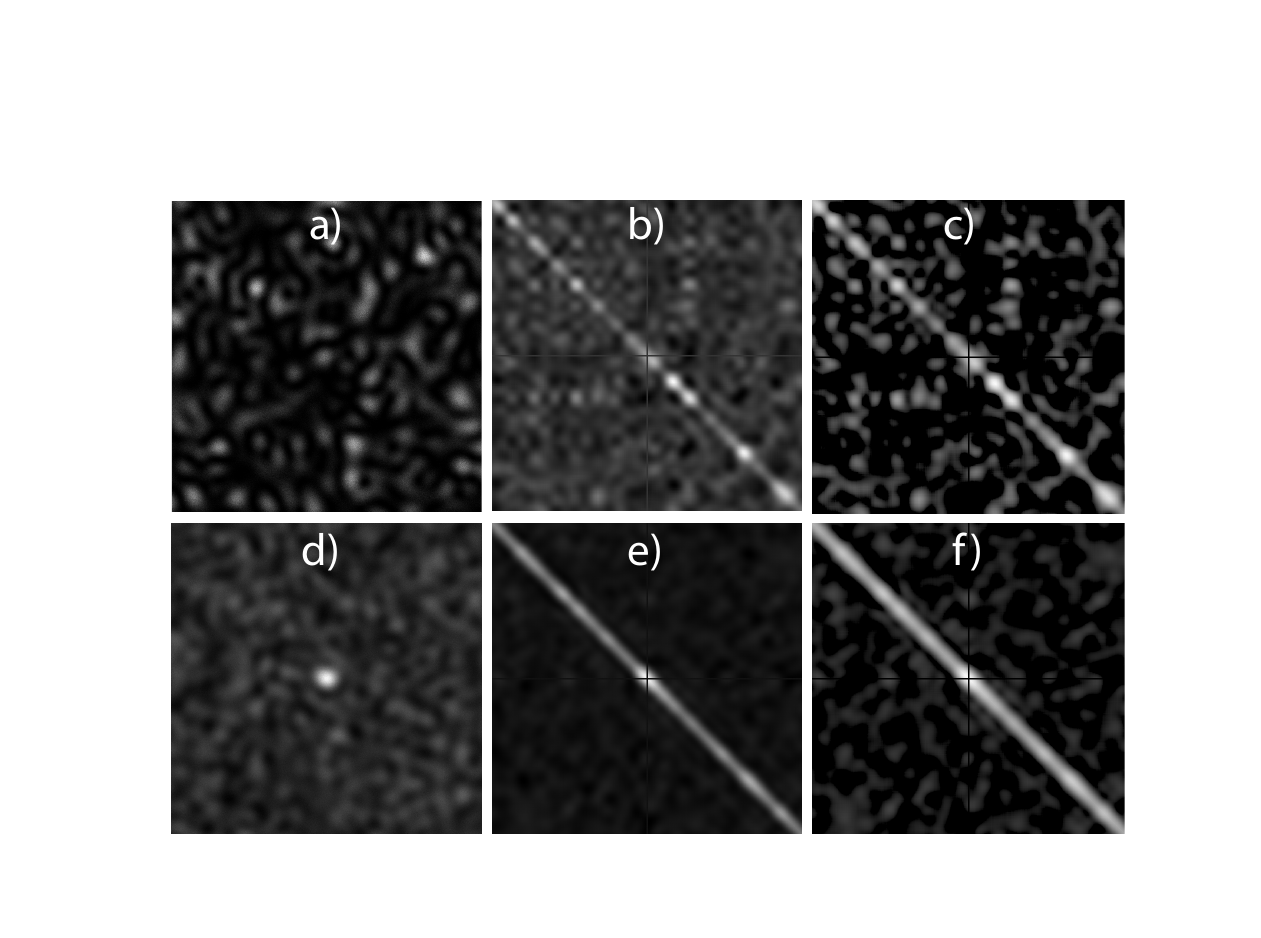}   
	\caption{
a) one simulated $G$\,=\,15 amplification event, as it would be registered by EMCCD camera (signal is supposed to be in the middle, but, in contrary to event depicted in Fig.~\ref{fig:images}, is not visible in this particular realization), b) covariance matrix of this event, c) square root of the covariance matrix (to reduce the contrast); d) the average of 20 simulated events, e) averaged covariances of 20 simulated events, f) square root of the averaged covariance matrix. Black crosshair denotes the centre of the average of covariance matrices.
	}
	\label{fig:averaged}
\end{figure}

We performed our numerical simulations using an iterative code prepared in \textit{Matlab} package. In order to test the usability of multimode OPA for high angular resolution astronomical imaging, we implemented the model described above. Note that Eq.~(\ref{Eq:ModesImagePlane}) is undefined, when both arguments are equal to zero. At the same time, point (0, 0) is important for the accuracy of the simulations, since in perfect situation (no noise) the centroid of the signal should be located there. According to \cite{badPixels}, biharmonic interpolation is the most efficient from all the interpolation methods used frequently in astronomy. Therefore, we fitted biharmonic surface to the surrounding of pixel corresponding to \,\mbox{$(\tilde{\rho},\tilde{\phi})$\,=\,0}\, and copied the appropriate value into this pixel. Since previously the value of this pixel was not defined, there is no possibility to estimate the goodness of such a fit at this pixel. The computed value is larger by 0.1985\% from the median of 8 nearest surrounding pixels (see Fig~\ref{fig:fitted}).

In order to simulate an individual image frame, a set of complex amplitudes $\alpha_{nm}$ is randomly chosen according to Eqs.~(\ref{Eq:palpha00}) and (\ref{Eq:palphanm}). For this realization, the intensity distribution in the image plane is given by:
\begin{equation}
	I(\tilde{\rho},\tilde{\phi}) = \Biggl|
	\sum_{{n=0} }^N
	\sum_m^n
	\alpha_{nm} u_{nm} (\tilde{\rho},\tilde{\phi}) \Biggr|^2,
\end{equation}
where, as previously, $m = -n, -n + 2, \ldots, n - 2,\, n$.
In practice, the summation over $n$ is truncated to a cut-off value $N = 100$ for which the included modes are effectively complete in the image area of interest. Numerically, this means that the sum $\sum_{nm} |u_{nm} (\tilde{\rho},\tilde{\phi})|^2$ with $n=0,1,\ldots, N$\ \ should be sufficiently close to one over the respective area. In our simulations we found that the deviation from one for the above sum is $\lesssim$10$^{\text{\,-11}}$. In the last step, the image plane is discretized into pixels and the number of photons in an individual pixel is chosen according to a Poissonian distribution with the mean equal to $I(\tilde{\rho},\tilde{\phi})$ times the pixel area.

The above procedure describes a simulation for a point source. For a non-point incoherent light source we needed in the first step to choose a single point emitting the radiation in the source plane and follow the procedure described above. The only modification is that the mode functions $u_{nm}(\tilde{\rho},\tilde{\phi})$ in the image plane introduced in Eq.~(\ref{Eq:ModesImagePlane}) should be displaced correspondingly to the location of the source. Care should be taken that the displaced mode functions also satisfy the effective completeness condition within the imaging area.

\vspace{2mm}
Let $r=\left|\alpha_{nm}\right|$ notes the amplitude of the single mode. Then Eqs. (\ref{Eq:palpha00}) and (\ref{Eq:palphanm}) can be rewritten in the polar coordinates as:
\begin{equation}
	p(r,\theta)=\frac{1}{\pi(G-1)}\exp\left(\frac{-r^2+2\sqrt{G}r\cos\theta-G}{G-1}\right)
\end{equation}
and
\begin{equation}
	p(r,\theta)=\frac{1}{\pi(G-1)}\exp\left(-\frac{r^2}{G-1}\right),
\end{equation}
respectively. We can marginalize the above equations over $\theta$ coordinate. In consequence, we get
\begin{equation}
	p(r)=\frac{2}{(G-1)}r\exp\left(\frac{r^2+G}{1-G}\right)I_0\left(\frac{2\sqrt{G}r}{G-1}\right)
	\label{probability1}
\end{equation}
and
\begin{equation}
	p(r)=\frac{2}{(G-1)}r\exp\left(\frac{r^2}{1-G}\right),
	\label{probability2}
\end{equation}
respectively. $I_i(x)$ is the modified Bessel function of the first kind. Let $r_0$ notes the amplitude of the signal. The probability that the amplitude of the single mode of the noise is bigger than $r_0$ is given by the formula:
\begin{equation}
	F(r_0)=\int_{r_0}^\infty p(r)dr=\exp\left(\frac{r_0^2}{1-G}\right),
\end{equation}
where $p(r)$ is given by Eq. (\ref{probability2}). The probability that the amplitude of a mode of the noise is bigger than the amplitude of the signal is expressed by:
\begin{equation}
	E = \int_0^\infty F(r_0) p(r_0)d r_0 = \frac{1}{2}\exp\left(\frac{G}{2(1-G)}\right), \label{eq:Er}
\end{equation}
where $p(r_0)$ is given by Eq. (\ref{probability1}).
From the above equation we can obtain the probability that the amplitude of any mode of noise is bigger than the amplitude of the signal:
\begin{align}
	P &=  \nonumber
	\sum_{i=1}^N(-1)^{i+1}{{N}\choose{i}}E^i\\ \nonumber
	&= \sum_{i=1}^N(-1)^{i+1}{{N}\choose{i}}\frac{1}{2^i}\exp\left(\frac{i G}{2(1-G)}\right)\\
	&= 1-\left(1-\frac{1}{2}\exp\left(\frac{G}{2(1-G)}\right)\right)^N,
	\label{probFinal}
\end{align}
where $N$ is a number of the modes of the noise.

\section{Results -- centroid search method}
\label{sec:res}

Fig.~\ref{fig:images}\, presents an exemplary outcome of the simulations: a) density of e-m wave in the image plane, b) surface distribution of the photons and c) an image as it would be registered by a modern EMCCD camera\footnote{For a review on EMCCD image formation, see~\cite{emccd}, sec. 2 therein.}, which introduces excess (multiplication) noise. An exemplary result depicted in Fig.~\ref{fig:images}c demonstrates that the camera efficiency (registration error) is not an essential issue here, as the image degradation after adding registration error (SSIM\footnote{Structural Similarity Index is a measure of image similarity. Its value ranges from -1 to 1 (identical images). For detail see \cite{ssim}.}\,=\,0.7060) is insignificant in terms of distinguishing of the clumps structures. For a comparison with the experiment, we can look at Fig.~2 from~\cite{clumps}, which presents experimentally acquired light speckles which are very similar to our simulated images (however, \citet{clumps} present the result of a single-mode light experiment, which is governed by Bose-Einstein distribution, therefore some deviations would be justified). The noise has a clumpy structure and the signal looks like one of the clumps, because the modes of both the signal and the clumpy structure are described by the same expression: Eq.~(\ref{Eq:ModesImagePlane}). The only difference between the signal and the clumps is given by the probability distribution of the amplitude of modes (see Eqs.~(\ref{Eq:palpha00}) and (\ref{Eq:palphanm})). Such clumps in the laser optics are called \textit{speckles}~\citep{speckle1, speckle2}.

In the MF-based signal analysis used in \cite{QT3}, the simulated image was first convolved with the Gaussian profile \mbox{($\sigma$ = 10)}, and then the centroid (centre of the signal photon cloud) was obtained from the position of the maximum value of such a filtered image. As in the outcome of Eq.~(\ref{Eq:ModesImagePlane}) in the present model both the signal and the noise have a form of in average identical clumps , there is no obvious reason MF would distinguish the signal from the noise, as it was the case in the previous models. The only way to recover the signal is to assume, that in some realizations the signal clump may be stronger than the noise clumps. This is a consequence of the probability distributions of the amplitude of the signal (Eq.~\ref{Eq:palpha00}) and clumps (Eq.~\ref{Eq:palphanm}). Therefore, we looked for the highest point in the image, as it was done in~\cite{QT3}.

In the exemplary realization of the density of the electromagnetic wave in the image plane depicted in~Fig.~\ref{fig:images} the position of the signal is clearly easy to recover, but in most cases the signal was not localized correctly. The probability that the amplitude of any mode of noise is bigger than the amplitude of the signal is given by formula (\ref{probFinal}). In Fig.~\ref{fig:averaged} the gain parameter $G$\,=\,15. Fig.~\ref{fig:averaged}a) presents an exemplary realization, where the signal is supposed to be located in the middle, but it is not visible. Fig.~\ref{fig:averaged}b) presents a covariance matrix of this event. Fig.~\ref{fig:averaged}c) presents the square root of the covariance matrix to reduce the contrast. Fig.~\ref{fig:averaged}d), e), and f) presents the analogical information, but for the average of 20 consecutive events. While it is not guaranteed, that it will be possible to localize the signal in any particular event, it is always possible in the average of large enough number of events. FWHM (Full Width at Half Maximum) sizes and intensities of the noise clumps are on average very similar across the considered field of view, which is depicted by the diagonal in Fig.~\ref{fig:averaged}b), c), e) and f). Because modes of the signal and the clumps are given by the same formula (Eq.~(\ref{Eq:ModesImagePlane})), the shape and size of noise clumps are indistinguishable from the signal of interest (clones cloud), which is best shown by the diagonal in Fig.~\ref{fig:averaged}f). Eq.~(\ref{probFinal}) describes statistically how often the signal steps out above the noise as a function of the gain $G$.

\subsection{Angular resolution -- extended source}
\label{sec:angRes}

\begin{figure}
\centering
	\includegraphics[width=\linewidth, trim= 2.3cm 13cm 2cm 12.5cm]{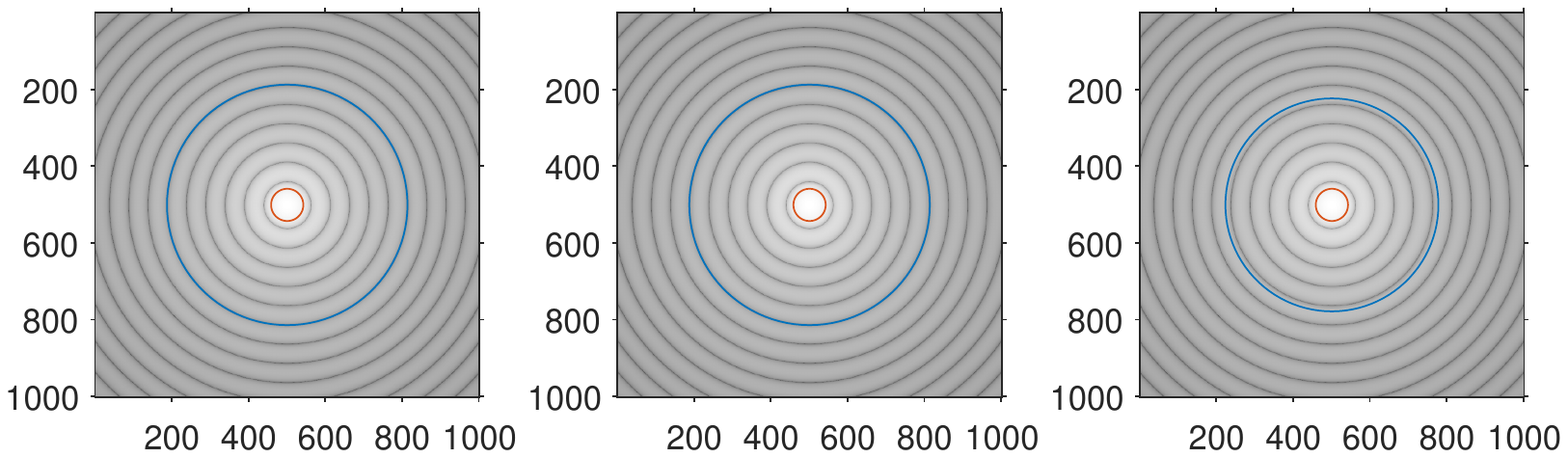}   
	\caption{
The exemplary results of centroid estimation error overplotted on the log intensity-scaled diffraction pattern. Left: unprocessed frame. Middle: blurred frame. Right: MF-filtered frame. The average error for QT is in blue, for CT -- in red.
	}
	\label{fig:exemplarResultExtended}
\end{figure}

To compare the efficiency of CT and QT in high angular resolution imaging of extended sources, we compared the "pencil size" of both of them. To simulate results with would be obtained by CT, we simulated classical optics image formation of a distant point source. In this process an Airy pattern is successively drawn by incoming photons. In the case of the CT, we distributed the signal according to the squared absolute values of $u_{0, 0}$, which is in fact a diffraction (Airy) pattern of the assumed optical system. In this distribution, an average Euclidean distance of a count from the centre of the image (depicted in Fig.~\ref{fig:exemplarResultExtended} with red circles) was 42.97 pixels. For reference, mean Euclidean distance of all pixels in the frame from the central pixel, computed from the equation:
\begin{equation}
	d_{all} =\frac{\sum_{x=1}^{1001}\sum_{y=1}^{1001} \sqrt{(x-501)^2+(y-501)^2} }{ 1001^2},
\label{eq:randomDist}
\end{equation}
was 383.98\,pix.

\begin{figure}
\centering
	\includegraphics[width=\linewidth, trim= 3.5cm 12.7cm 4.2cm 12.5cm]{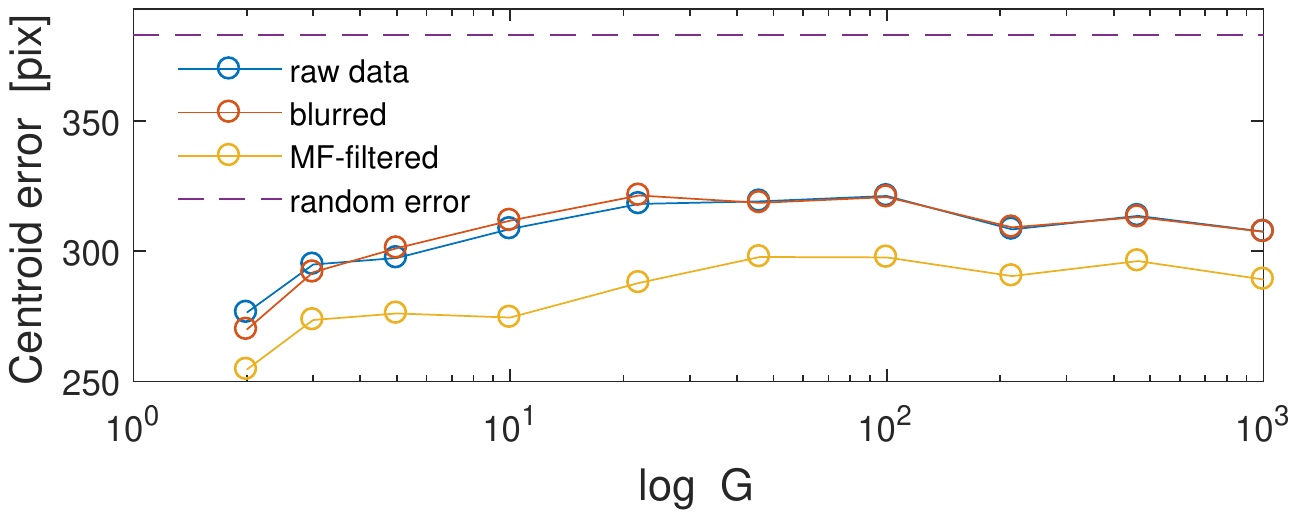}   
	\caption{
Centroid estimation error of the QT in high angular resolution extended source imaging. All three types of tests are presented. Each point represents a mean of 1k iterations. Average random pixel error is overplotted by dashed line.
	}
	\label{fig:resultQTextended}
\end{figure}

\begin{figure*}
\centering
	\includegraphics[width=0.7\linewidth, trim= 3cm 8.5cm 3cm 8.2cm]{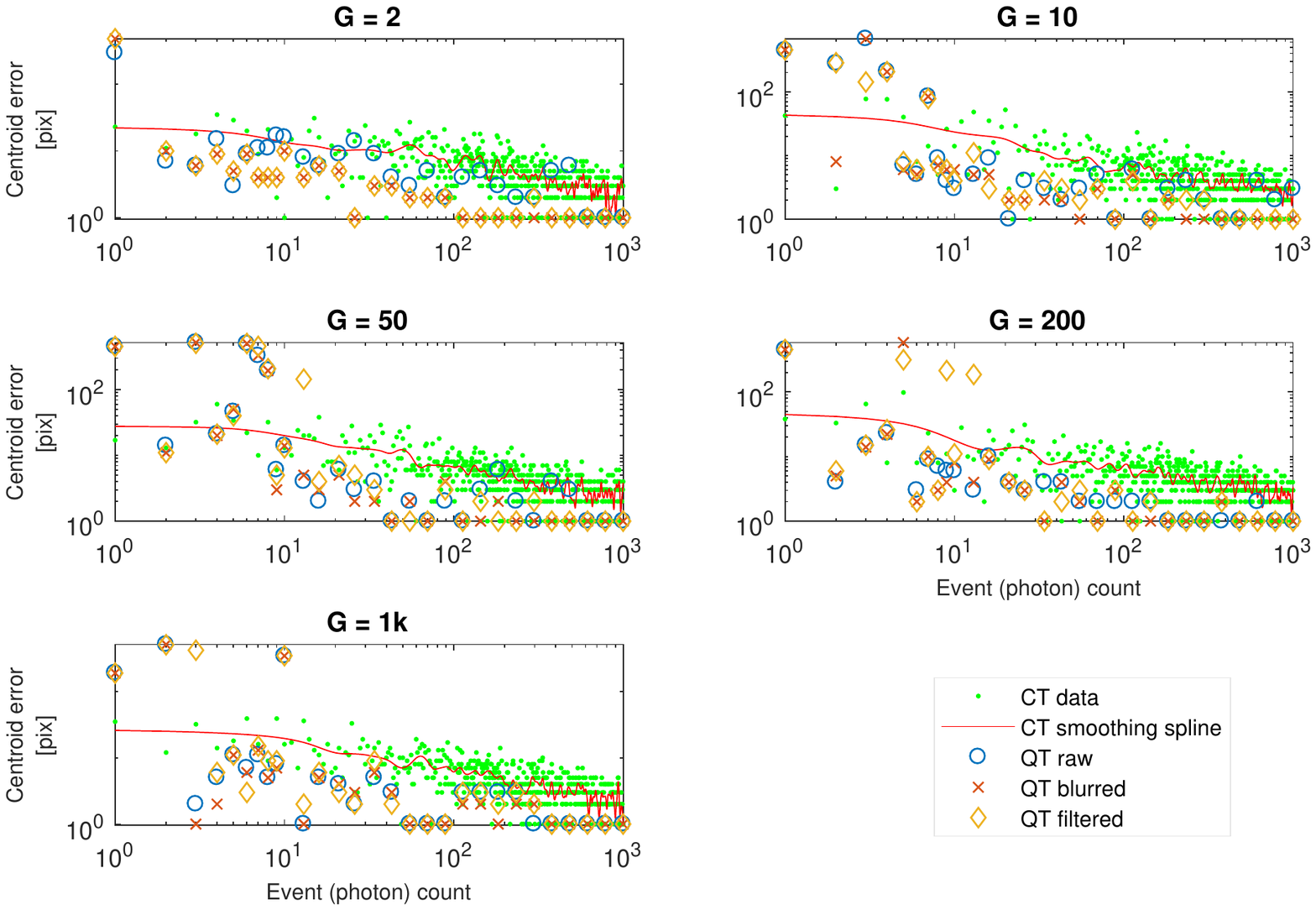}   
	\caption{
Comparison of centroid errors produced by CT and QT in localization of a point source. Green dots denote the efficiency of classic optics. For each $G$ the results for CT were computed anew. Circles, crosses and diamonds denote the QT/OPA efficiency.
	}
	\label{fig:diagnostics}
\end{figure*}

In the case of the QT, after each amplification event we computed the euclidean distance of the highest peak from the known true source position. We performed three groups of tests. In the first one, we did not modify the frame in any way. In the second one, we convolved the frame with a Gaussian spanning on 5$\times$5 pixels and \mbox{$\sigma$ = 0.5}. That was done in order to slightly blur the image and thereby reduce the influence of the shot noise on the centroid estimation. In the third test we performed MF on the averaged frame using the diffraction pattern as a known template. This is in essence equivalent to the test 2, but instead of an arbitrary assumed Gaussian, we used a diffraction pattern in the convolution. The results varied depending on the used amplification gain $G$ and the signal analysis method. An example depicting the procedure is presented in Fig.~\ref{fig:exemplarResultExtended}. In this example, the mean localization error ("pencil size") in the case of QT varies from 313.32 ($\sigma$ = 190.31) pix. in both the unprocessed and blurred frames to 277.47 ($\sigma$ = 180.34) in MF-filtered frame. All test were done using 1001$\times$1001 frames.

Fig.~\ref{fig:resultQTextended} presents the results for all three tests. For every tested $G$, the centroid error of the QT is many times below the one of CT ($\sim$43\,pix). But it is still better, than an average distance from the centre of random pixel within the frame (Eq.~(\ref{eq:randomDist})), so in the case of all $G$ statistically there is a weak correlation between highest peak and a true position of the source.

In sum, in the framework of the model presented here, the noise has a correlated space-dependent structure, which was not assumed in previous models \citep{QT3}. Therefore, there is no obvious way to localize the signal, since in most cases the highest peak appears anywhere in the frame at a position too weakly correlated to the position of the signal (see Eq.~(\ref{probFinal})), so at this step the procedure fails. To obtain any gain in the resolution, the signal localization error for every event would have to be in average lower than the Airy disc FWHM. It implies that the use of centroid search method for signal analysis from OPA in astronomy does not allow for a resolution gain in imaging of extended sources.

\subsection{Localization of a distant point source}
\label{sec:loc}

Another possible use of the QT are astrometric measurements of the position of a distant point source. As before, we performed our simulations using \mbox{1001$\times$1001} pixel images. The count of photons constituting an undersampled Airy pattern was equal to the amplification event count in the QT. For CT we computed the dot product centroid of the signal and the Euclidean distance of the centroid to the centre of the diffraction pattern of the source, of which the position was estimated. In the case of the QT, we searched for the highest point in the averaged set of images of the outcome of the amplification process degraded by the shot noise. As in the case of extended source, we used three methods of signal analysis for QT: an analysis of (a) raw data, (b) blurred data and (c) match-filtered data.

According to our tests, for the same number of events, the localization error of QT is in average lower than that of the classical optics. Fig.~\ref{fig:diagnostics} presents exemplary result for $G$\,=\,2, 10, 50, 200 and 1k of the centroid estimation error as a function of photons (in the case of CT) and amplification events (in the case of QT) count. The tests based on processed QT data (blurred and MF-filtered) shows lower error than test based on raw data. For relatively small number of events (10-20) the supremacy of QT over CT is unstable and gets stabilized for larger iteration counts. The exact result (Fig.~\ref{fig:diagnostics}) is a nonlinear function of: gain $G$ (weak dependence), photons/events count (strong dependence) and method of centroid computation (weak dependence), but, as for a rule of thumb, the Euclidean distance error tends to be $\sim$3$\times$ smaller with the use of OPA, if the event number exceeds 10-20.

The superiority of OPA probably originates from the fact that, beginning from considerably small numbers of events, the signal in an averaged frame is more intense than the noise clumps, thus providing an accurate information on the position of the source. At the same time, in the case of CT the Airy pattern is still not fully drawn, so the shot noise still significantly increases the localization error.

\section{Results -- speckle maxima method}
\label{sec:REVspceMax}
\begin{figure}
\centering
	\includegraphics[width=\linewidth, trim= 0.5cm 13cm 0.6cm 3.5cm]{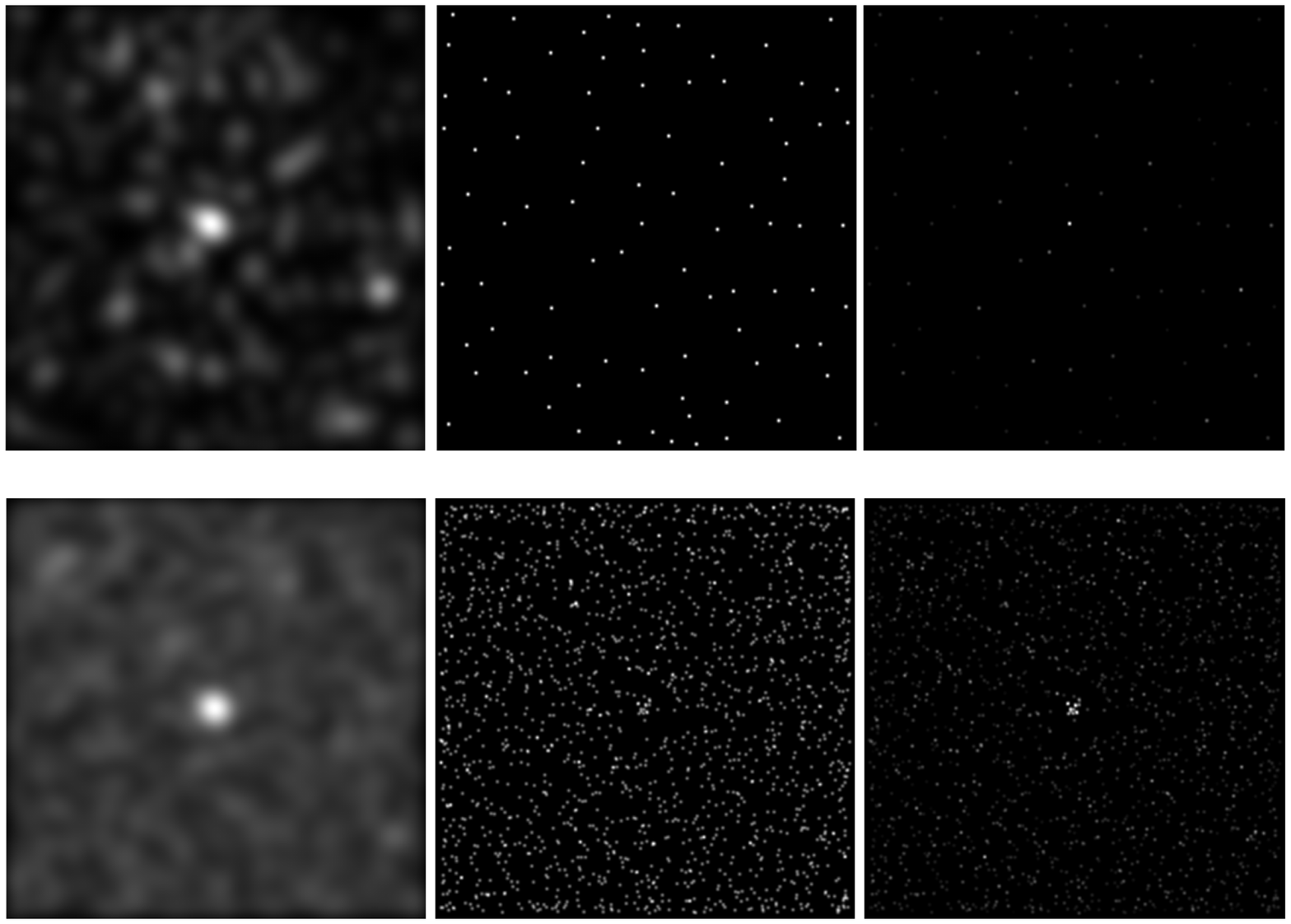}   
	\caption{
Localization of maxima of the speckles. 1$^{\text{st}}$ row: 1 iteration. 2$^{\text{nd}}$ row: stacks of 20 iterations. Left column: OPA output. Middle: localized maxima of the speckles (0-1 logic). Right: localized maxima of the speckles -- every maximum is represented by its value. Note that maxima of elongated speckles are also localized correctly. Gain $G$ = 10.
	}
	\label{fig:localization}
\end{figure}
Another method of signal analysis we tested is based on localization of the maxima of all the speckles. After acquiring each frame of 1001$\times$1001 pixels, we perform MF to remove the effects of the shot noise. As the next step, we analyze the frame using a 3$\times$3 pixels sliding window. To avoid boundary conditions problems, we run the sliding window from pixel (2, 2) and end it at (end-1, end-1). This is justified, as the signal is expected in the middle of the frame. We check if the pixel in the middle of the window has a higher value than any other pixel in the window. If it were the case, the pixel would be marked as a peak of a speckle. This way we localize maxima of all the speckles present on the frame (Fig.~\ref{fig:localization}). We verified this approach by carefully investigating the localized positions and no evidently false detections were found. It implies that the MF is able to cancel out efficiently the shot noise effects. The only questionable detections are localized near the edges of the frame, but they do not influence on the results, since, as mentioned, the signal is expected in the middle of the frame -- which is far enough from the edges (e.g. Fig~\ref{fig:localization}, left).

\subsection{Angular resolution -- extended source}
\label{sec:REVangRes}
\begin{figure}
\centering
	\includegraphics[width=0.95\linewidth, trim= 3.5cm 11.4cm 4.5cm 11.9cm]{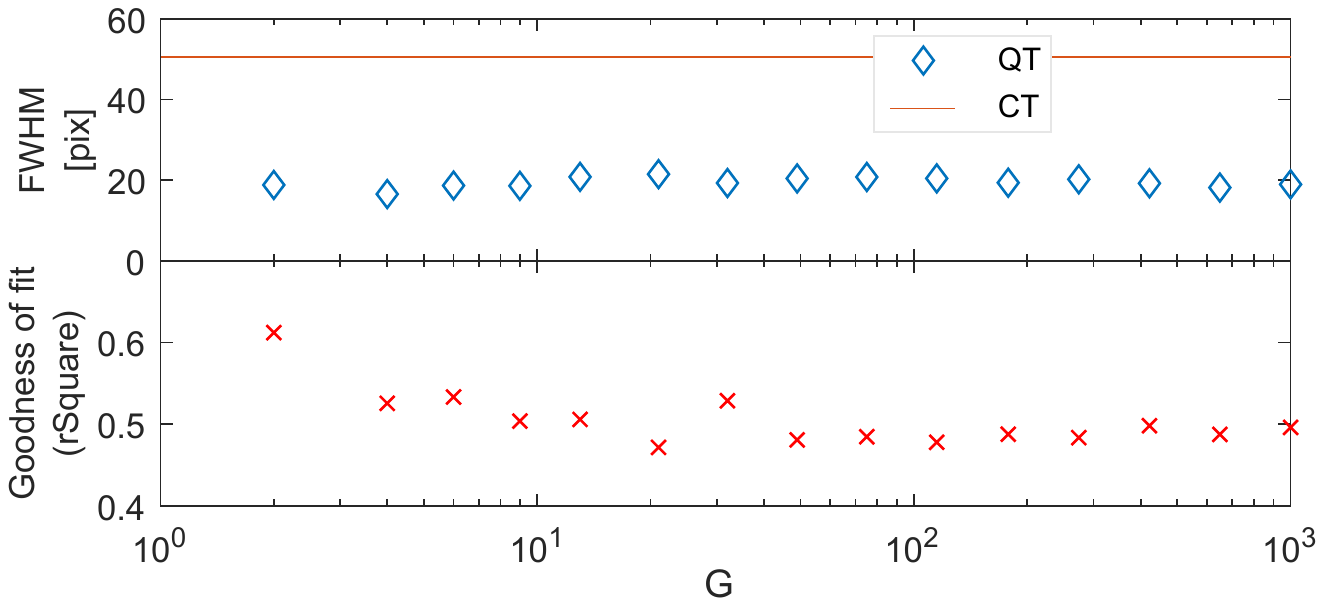}   
	\caption{
Top: comparison of the FWHM for QT and CT. Bottom: goodness of fit for QT (r-squared).
	}
	\label{fig:REVangResresults}
\end{figure}

After stacking a large enough number of localization results, a Gaussian-like shape is visible in the middle of the stack (Fig.~\ref{fig:localization}, bottom right). In the current method we assume it originates from the signal clumps and that it can be fitted by a 2D Gaussian surface. To ensure stable fit results, we do this using a stack of 2000 frames. We measure the FWHM of such a fitted Gaussian and compare it to the FWHM of $u_{0, 0}$. The results for a series of gain $G$ values are shown in Fig.~\ref{fig:REVangResresults}. The FWHM obtained with the use of OPA (mean 19.45\,pix, $\sigma$\,=\,1.27\,pix) is significantly narrower than the one obtained with the use of classical optics (50.45\,pix). It confirms that OPA is able to produce a gain in the resolution of 2D imaging.

\subsection{Localization of a distant point source}
\label{sec:REVloc}
\begin{figure}
\centering
	\includegraphics[width=\linewidth, trim= 2.5cm 9cm 3.5cm 8cm]{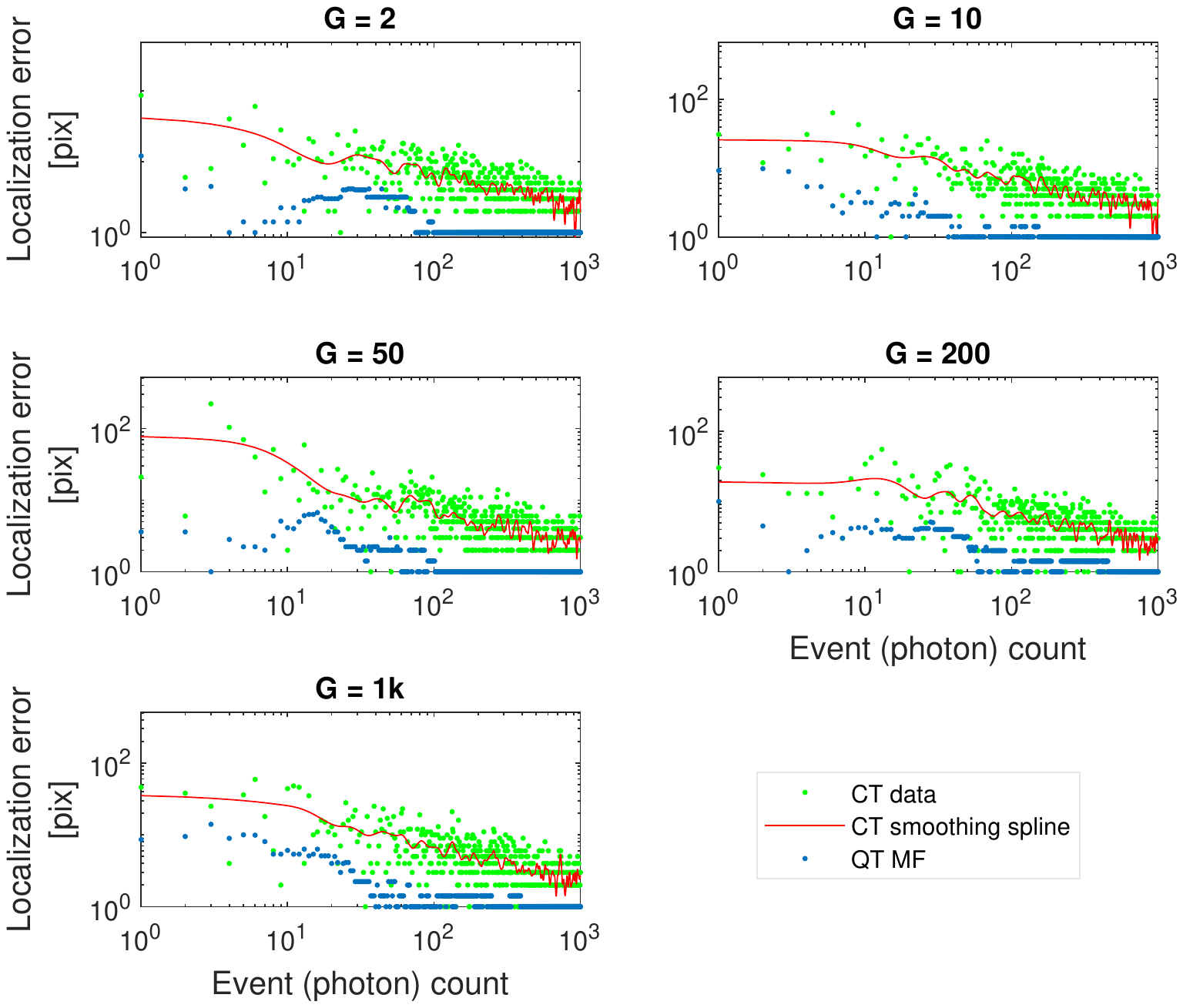}   
	\caption{
Comparison of localization error of a point source using speckle maxima method. Green dots denote the efficiency of classic optics. For each
G the results for CT were computed anew. Blue dots denote the QT/OPA efficiency.
	}
	\label{fig:REVpointSrc}
\end{figure}

As in Sec.~\ref{sec:loc}, we checked the efficiency for 1$\sim$1k events and a series of gain $G$ values. After acquiring a currently targeted number of localization-frames (e.g. Fig.~\ref{fig:localization}, bottom right), we stacked them and performed MF using $u_{0, 0}$ as a convolution kernel. We computed the Euclidean distance from the highest point of the convolution results to the middle of the frame and legitimized this distance as a point-source localization error. The results are presented in Fig.~\ref{fig:REVpointSrc}. It can be seen that OPA is able to achieve the precision limit of such a test more that an order of magnitude faster than CT.

\section{Conclusions}
\label{sec:Conclusions}
Our semi-classical, more accurate then previously considered model shows that the intrinsic parametric amplification noise tends to form random clumps when passing the telescope optics. This effect was not included in the previous models of the QT~\citep{QT1, QT2, errata, QT3, QT4} based on the first quantization. However, such clumps, called light speckles, are shown experimentally to exist~\citep{clumps}. The presence of these clumps means that it is difficult to distinguish the signal from the noise and implies that a different signal analysis procedure should be applied. Further modeling and analyzes of such a noise in greater details are needed, because an understanding of this issue is essential for practical realization of a Quantum Telescope.

According to our new results, the centroid search approach will not produce any considerable resolution gain. But the approach based on localizing maxima of the speckles -- the second signal analysis method we tested -- offers about threefold improvement over classical optics. Moreover, both signal analysis methods were shown to be more efficient in localization of a point source than the classical optics, given the same photon count per a measurement. Here again the speckle maxima method is superior to the centroid search approach.

In principle, it is possible to further increase the performance with the use of a noiseless amplification instead of the OPA. However, this approach was never tested in astronomy~\citep{noiseless1, noiseless2, noiseless3}. Recently, there appeared also other competitive to OPA techniques of overcoming the diffraction limit in far-field imaging~\citep{Chrostowski2017}. These possibilities are currently under investigation by our group.

\section*{Acknowledgments}
\label{sec:ack}

The authors thank our colleague, Dr. R. Demkowicz-Dobrzanski for useful and inspiring discussions and Dr. A. Popowicz for valuable suggestions concerning signal analysis. We also thank anonymous Referee for useful and constructive comments which allowed to improve the manuscript significantly.

\vspace{1mm}
{\bf Funding}\quad Polish National Science Centre grant no 2016/21/N/ST9/00375. European Commission FP7 projects SIQS (Grant Agreement No. 600645) and PhoQuS@UW (Grant Agreement No. 316244) co-financed by the Polish Ministry of Science and Higher Education. MNSW: 71501E-338/M/2017. Foundation for Polish Science TEAM project ``Quantum Optical Communication Systems'' co-financed by the European Union under the European Regional Development Fund.

\bibliographystyle{mnras}
\bibliography{biblio}

\bsp	
\label{lastpage}
\end{document}